\documentclass{cernyrep}

\usepackage{endnotes}
\usepackage{color}
\newcommand{\aq}[1]{}
\newcommand{\tq}[1]{}
\newcommand{\ced}[1]{#1}

\usepackage{amssymb}
\usepackage[final]{epsfig}
\usepackage{color}
\usepackage{graphicx}
\usepackage{lineno}
\begin{document}



\title{Particle Physics Instrumentation}

\author{Werner Riegler
}
\institute{CERN, Geneva, Switzerland}

\maketitle

\begin{abstract}
 This report summarizes a series of three lectures aimed at giving an overview of basic particle detection principles, the interaction of particles with matter, the application of these principles in modern detector systems, as well techniques to read out detector signals in high-rate experiments.
\end{abstract}



\section{Introduction}

``New directions in science are launched by new tools much more often than by new concepts'' is a famous quote from Freeman Dyson's book \textit{Imagined Worlds}. This is certainly true for the field of particle physics, where new tools such as the cloud chamber, bubble chamber, wire chamber, solid-state detectors, accelerators, etc. have allowed physicists to enter into unchartered territory and to discover unexpected phenomena, the understanding of which has provided a deeper insight into the nature of matter. Looking at all Nobel Prize winners connected to the Standard Model of particle physics, one finds many more experimentalists and ``instrumentalists'' than theoretically orientated physicists, which is a strong indicator of the essence of new tools for advancing our knowledge.

This report will first discuss a few detector systems in order to illustrate the detector needs and specifications of modern particle physics experiments. Then the interaction of particles with matter, which is of course at the heart of particle detection, will be reviewed. Techniques for tracking with gas detectors and solid-state detectors as well as energy measurement with calorimeters are then elaborated. Finally, the tricks on how to process the signals from these detectors in modern high-rate applications will be discussed.

\section{Examples of detector systems}

The \ced{Large Hadron Collider (LHC)} experiments ATLAS, CMS, ALICE and LHCb are \ced{currently} some of the most prominent detectors because of their size, complexity and rate capability. Huge magnet systems, which are used to bend the charged particles in order to measure their momenta, dominate the mechanical structures of these experiments. Proton collision rates of 1~GHz, producing particles and jets of TeV-scale energy, \ced{present} severe demands in terms of spectrometer and calorimeter size, rate capability and radiation resistance. The fact that only about 100 of the $10^9$ events per second can be written to disk \ced{necessitates} highly complex online event selection, i.e.\ ``triggering'. The basic layout of these collider experiments is quite similar. Close to the interaction point there are several layers of pixel detectors that allow \ced{the collision vertices to be distinguished and measured} with precision on the tens of micrometres level. This also allows short-lived B and D mesons \ced{to be identified} by their displaced decay vertices. In order to follow the tracks along their curved path up to the calorimeter, a few metres distant from the collision point, one typically uses silicon strip detectors or gas detectors at larger radii. CMS has an ``all-silicon tracker'' up to the calorimeter, while the other experiments use also gas detectors like so-called straw tubes or a time projection chamber. The trackers are then followed by the electromagnetic and hadron calorimeter, which measures the energy of electrons, photons and hadrons by completely absorbing them in very large amounts of material. The muons, the only particles able to pass through the calorimeters, are then measured at even larger radii by dedicated muon systems. The sequence of vertex detector, tracker for momentum spectrometry, calorimeter for energy measurement followed again by tracking for muons is the classic basic geometry that underlies most collider and even fixed-target experiments. It allows one to distinguish electrons, photons, hadrons and muons and to measure their momenta and energies.

The ALICE and LHCb experiments use a few additional detector systems that allow  \ced{different hadrons to be distinguished}. By measuring the particle's velocity in addition to the momentum, one can identify the mass and therefore the type of hadron. This velocity can be determined by measuring time of flight, the Cerenkov angle or the particle's energy loss. ALICE uses, in addition, the transition radiation effect to separate electrons from hadrons, and has therefore implemented almost all known tricks for particle identification. Another particle detector using all these well-established techniques is the Alpha Magnetic Spectrometer (AMS) that has recently been installed on the \textit{International Space Station}. It is aimed at measuring the primary cosmic-ray composition and energy distribution.

More ``exotic'' detector geometries are used for neutrino experiments, which demand huge detector masses in order to make the neutrinos interact. The IceCube experiment at the South Pole uses one cubic kilometre of ice as the neutrino detection medium to look for neutrino point sources in the Universe. Neutrinos passing \ced{through} the Earth from the Northern Hemisphere interact deep down under the ice and the resulting charged particles are travelling upwards at speeds larger than the speed of light in the ice. They therefore produce Cerenkov radiation, which is detected by a series of more than 5000 photon detectors that are immersed into the ice and look downwards. An example of an accelerator-based neutrino experiment is the CERN Neutrino to Gran Sasso (CNGS) beam. A neutrino beam is sent from CERN over a distance of 732~km to the Gran Sasso laboratory in Italy, where some large neutrino detectors are set up. One of them, the OPERA detector, uses more than 150\,000 lead bricks as neutrino target. The bricks are built up from alternating sheets of lead and photographic emulsion, which allows tracking with \ced{the micrometre precision necessary} to identify the tau leptons that are being produced by interaction of tau neutrinos. This ``passive'' detector is followed by trigger and tracking devices, which detect secondary particles from the neutrino interactions in the lead bricks and identify the bricks where an interesting event has taken place. To analyse the event, the bricks have then to be removed from the assembly and the photographic emulsion must be developed.

These are only a few examples from a large variety of existing detector systems. It is, however, important to bear in mind that there are only a few basic principles of particle interaction with matter that underly all these different detectors. It is therefore worth going through them in detail.

\section{Basics of particle detection}

The Standard Model of particle physics counts 17 particles, namely six quarks, six leptons, photon, gluon, W and Z bosons, and the hypothetical Higgs particle. Quarks, however, are not seen as free particles; rather, they combine into baryons and mesons, of which there are hundreds. How can we therefore distinguish all these different particle types in our detectors? The important fact is that, out of the hundreds of known hadrons, only 27 have a lifetime that is long enough such that they can leave a track \ced{$>1~\mu\mathrm{m}$} in the detector. All the others decay ``on the spot'' and can only be identified and reconstructed through kinematic relations of their decay products like the ``invariant mass''. Out of these 27 particles, 13 have lifetimes that make them decay after a distance between a few hundred micrometres and a few millimetres at GeV energies, so they can be identified by their decay vertices, which are only a short distance from the primary collision vertex (secondary vertex tagging). The 14 remaining particles are the only ones that can actually ``fly'' though the entire detector, and the following eight are by far the most frequent ones: electron, muon, photon, charged pion, charged kaon, neutral kaon, proton and neutron. The principle task of a particle detector is therefore to identify and measure the energies and momenta of these eight particles.

Their differences in mass, charge and type of interaction are the key to their identification, which will be discussed in detail later.
The electron leaves a track in the tracking detector and produces a shower in the electromagnetic (EM) calorimeter. The photon does not leave a track but also produces a shower in the EM calorimeter. The charged pion, charged kaon and the proton show up in the tracker but \ced{pass through} the EM calorimeter and produce hadron showers in the hadron calorimeter. The neutral kaon and the neutron do not show tracks and shower in the hadron calorimeter. The muon is the only particle than manages to pass through even the hadron calorimeter and is identified by tracking detectors behind the calorimeters. How to distinguish between pion, kaon and proton is typically the task of specific particle identification (PID) detectors.

\section{Interaction of particles with matter}

The processes leading to signals in particle detectors are now quite well understood and, as a result of available computing power and simulation programs like GEANT or GARFIELD, one can simulate detector responses to the level of a few percent based on fundamental microphysics processes (atomic and nuclear cross-sections). By knowing the basic principles and performing some ``back-of-the-envelope calculations'', it is possible to estimate detector response to the 20--30\% level.

It sounds obvious that any device that is to detect a particle must interact with it in some way. In accelerator experiments, however, there is  a way to detect neutrinos even if they do not interact in the detector. Since the total momentum of the colliding particles is known, the sum of all momenta of the produced particles must amount to the same number, owing to momentum conservation. If one uses a hermetic detector, the measurement of missing momentum can therefore be used to detect the momentum vector of the neutrino!

The electromagnetic interaction of charged particles with matter lies at the heart of all particle detection. We can distinguish six types of these interactions: atomic excitation, atomic ionization, bremsstrahlung, multiple scattering, Cerenkov radiation and transition radiation. We will discuss them in more detail in the following.

\subsection{Ionization and excitation}

A charged particle \ced{passing through} an atom will interact through the Coulomb force with the atomic electrons and the nucleus. The energy transferred to the electrons is about 4000 times larger compared to the energy transferred to the nucleus because of the much higher mass of the nucleus. We can therefore assume that energy is transferred only to the electrons. In a distant encounter between a passing particle and an electron, the energy transfer will be small -- the electron will not be liberated from the atom but will just go to an excited state. In a close encounter the energy transfer can be large enough to exceed the binding energy -- the atom is ionized and the electron is liberated. The photons resulting from de-excitation of the atoms and the ionization electrons and ions are used in particle detectors to generate signals that can be read out with appropriate readout electronics.

The faster the particle is \ced{passing through} the material, the less time there is for the Coulomb force to act, and the energy transfer for the non-relativistic regime therefore decreases with particle velocity $v$ as $1/v^2$. If the particle velocity reaches the speed of light, this decrease should stop and stay at a minimum plateau. After a minimum for Lorentz factors $\gamma = 1 / \ced{ \sqrt{1-v^2/c^2} } $ of $\approx 3$,
however, the energy loss increases again because the kinematically allowed maximum energy that can be transferred from the incoming particle to the atomic electron is increasing. This rise goes with \ced{$\log\gamma$} and is therefore called the relativistic rise. Bethe and Bloch devised a quantum-mechanical calculation of this energy loss in the 1930s. For ultra-relativistic particles, the very strong transverse field will polarize the material and the energy loss will be slightly reduced.

The energy loss is, in addition, independent of the mass of the incoming particle. Dividing the energy loss by the density of the material, it becomes an almost universal curve for all materials. The energy loss of a particle with $\gamma \approx 3$ is around $1\mbox{--}2 \times \rho[\mathrm{g/cm}^{3}]$~MeV/cm. Taking iron as an example, the energy  for a high-energy particle due to ionization and excitation is about 1~GeV/m. The energy loss is also proportional to the square of the particle charge, so a helium nucleus will deposit four times more energy compared to a proton of the same velocity.

Dividing this energy loss by the ionization energy of the material, we can get a good estimate of the number of electrons and ions that are produced in the material along the track of the passing particle. Since the energy deposited is a function of the particle's velocity only, we can use it to identify particles: measuring the momentum by the bending in a magnetic field and the velocity from the energy loss, we can determine the mass of the particle in certain momentum regions.

If a particle is stopped in a material, the fact that the energy loss of charged particles increases for smaller velocities results in large energy deposits at the end of the particle track. This is the basis of hadron therapy, where charged particles are used for tumour treatment.  These particles deposit a large amount of dose inside the body at the \ced{location} of the  tumour without exposing the \ced{overlying} tissue to high radiation loads.

This energy loss is, of course, a statistical process, so the actual energy loss will show fluctuations around the average given by the Bethe--Bloch description. This energy-loss distribution was first described by Landau and it shows a quite asymmetric tail towards large values of the energy loss. This large fluctuation of the energy loss is one of the important limiting factors of tracking detector resolution.

\subsection{Multiple scattering, bremsstrahlung and pair production}

The Coulomb interaction of an incoming particle with the atomic nuclei of the detector material results in deflection of the particle, which is called multiple scattering. A particle entering a piece of material perpendicular to the surface will therefore have a probability of exiting at a different angle, which has a Gaussian distribution with a standard deviation that depends on the particle's properties and the material. This standard deviation is inversely proportional to the particle velocity and the particle momentum, so evidently the effect of multiple scattering and related loss of tracking resolution and therefore momentum resolution is worst for low-energy particles. The standard deviation of the angular deflection is, in addition, proportional to the square root of the material thickness, so clearly one wants to use the thinnest possible tracking devices. The material properties are summarized in the so-called radiation length $X_0$, and the standard deviation depends on the inverse root of that. Materials with small radiation length are therefore not well suited to the volume of tracking devices. This radiation length $X_0$ is proportional to $A/\rho Z^2$ where $A$, $\rho$ and $Z$ are the nuclear number, density and atomic number of the material. Tracking systems therefore favour materials with very low atomic number like beryllium for beampipes, carbon fibre and aluminium for support structures, and thin silicon detectors or gas detectors as tracking elements.

The deflection of the charged particle by the nuclei results in acceleration and therefore emission of electromagnetic radiation. This effect is called ``bremsstrahlung'' and it plays a key role in calorimetric measurements. The energy loss of a particle due to bremsstrahlung is proportional to the particle energy and inversely proportional to the square of the particle mass. Since electrons and positrons are very light, they are the only particles where energy loss due to bremsstrahlung can dominate over energy loss due to ionization at typical present accelerator energies. The energy of a high-energy electron or positron travelling a distance $x$ in a material decreases as $\exp(-x/X_0)$, where $X_0$ is again the above-mentioned radiation length. The muon, the next lightest particle, has about 200 times the electron mass, so the energy loss from bremsstrahlung is 40\,000 times smaller at a given particle energy. A muon must therefore have an energy of more than 400~GeV in order to have an energy loss from bremsstrahlung that dominates over the ionization loss. This fact can be used to distinguish them from other particles, and it is at the basis of electromagnetic calorimetry through a related effect, the so-called pair production.

A high-energy photon has a certain probability of converting into an electron--positron pair in the vicinity of a nucleus. This effect is closely related to bremsstrahlung. The average distance that a high-energy photon travels in a material before converting into an electron--positron pair is also approximately given by the radiation length $X_0$. The \ced{alternating}\aq{Is ``alternating'' correct -- or alternative?} processes of bremsstrahlung and pair production result in an electromagnetic cascade (shower) of more and more  electrons and positrons with increasingly degraded energy until they are stopped in the material by ionization energy loss. We will come back to this in the discussion of calorimetry.

\subsection{Cerenkov radiation}

Charged particles \ced{passing through} material at velocities larger than the speed of light in the material produce an electromagnetic shock wave that materializes as electromagnetic radiation in the visible and ultraviolet range, the so-called Cerenkov radiation. With $n$ being the refractive index of the material, the speed of light in the material is $c/n$, so the fact that a particle does or does not produce Cerenkov radiation can be used to apply a threshold to its velocity. This radiation is emitted at a characteristic angle with respect to particle direction. This Cerenkov angle $\Theta_c$ is related to the particle velocity $v$ by $\cos \Theta_c = c/nv$, so by measuring this angle, one can determine the velocity of a charged particle.

\subsection{Transition radiation}

Transition radiation is emitted when a charged particle crosses the boundary between two materials of different permittivity. The probability of emission is proportional to the Lorentz factor $\gamma$ of the particle and is only appreciable for ultra-relativistic particles, so it is mainly used to distinguish electrons from other hadrons. As an example a particle with $\gamma = 1000$ has a probability of about 1\% to emit a photon on the transition between two materials, so one has to place many layers of material in the form of sheets, foam or fibres in order to produce a measurable amount of radiation. The energy of the emitted photons is in the keV region, so the fact that a charged particle is accompanied by X-rays is used to identify it as an electron or positron.

\section{Detector principles}

In the previous section we have seen how charged particles leave a trail of excited atoms and electron--ion pairs along their track. Now we can discuss how this is used to detect and measure them. We will first discuss detectors based on atomic excitation, so-called scintillators, where the de-excitation produces photons, which are reflected to appropriate photon detectors. Then we discuss gaseous and solid-state detectors based on ionization, where the electrons and ions (holes) drift in electric fields, which induces signals on metallic readout electrodes connected to readout electronics.

\subsection{Detectors based on scintillation}

The light resulting from complex de-excitation processes is typically in the ultraviolet to visible range. The three important classes of scintillators are the noble gases, inorganic crystals and polycyclic hydrocarbons (plastics). The noble gases show scintillation even in their liquid phase. An application of this effect is the liquid argon time projection chamber where the instantaneous light resulting from the passage of the particle can be used to mark the start signal for the drift-time measurement. Inorganic crystals show the largest light yield and are therefore used for precision energy measurement in calorimetry applications and also in nuclear medicine. Plastics constitute the most important class of scintillators owing to their cheap industrial production, robustness and mechanical stability. The light yield of scintillators is typically a few percent of the energy loss. In 1~cm of plastic scintillator, a high-energy particle typically loses 1.5~MeV, \ced{of which 15~keV goes into visible light, resulting in} about 15\,000 photons. In addition to the light yield, the decay time, i.e.\ the de-excitation time, is an important parameter of the scintillator. Many inorganic crystals such as NaI or CsI show very good light yield, but have decay times of tens, even hundreds, of nanoseconds, so they have to be carefully chosen considering the rate requirements of the experiments. Plastic scintillators, on the other hand, are very fast and \ced{have decay times on only the nanosecond scale}, and they are therefore often used for precision timing and triggering purposes.

The photons produced inside a scintillator are internally reflected to the sides of the material, where so-called ``light guides'' are attached to guide the photons to appropriate photon detection devices. A very efficient way \ced{to extract the light is to use} so-called wavelength shifting fibres, which are attached to the side of the scintillator materials. The light entering the fibre from the scintillator is converted into a longer wavelength there and it can therefore not reflect back into the scintillator. The light stays in the fibre and is internally reflected to the end, where again the photon detector is placed.

The classic device used to convert these photons into electrical signals is the so-called photomultiplier. A photon hits a photocathode, a material with very small work function, and an electron is liberated. This electron is accelerated in a strong electric field to a dynode, which is made from a material with high secondary electron yield. The one electron hitting the surface will therefore create several electrons, which are again guided to the next dynode, and so on, so that out of the single initial electron one ends up with a sizeable signal of, for example, $10^7$--$10^8$ electrons.

In recent years, the use of solid-state photomultipliers, the so-called avalanche photodiodes (APDs), has become very popular, owing to their much lower price and insensitivity to magnetic fields.

\subsection{Gaseous detectors}

A high-energy particle leaves about 80 electron--ion pairs in 1~cm of argon, which is not enough charge to be detected above the readout electronics noise of typically a few hundred to a few thousand electrons, depending on the detector capacitance and electronics design. A sizeable signal is only seen if a few tens or hundreds of particles cross the gas volume at the same time, and in this operational mode such a gas detector, consisting of two parallel metal electrodes with a potential applied to one of them, is called an ``ionization chamber''. In order to be sensitive to single particles, a gas detector must have internal electron multiplication. This is accomplished most easily in the wire chamber. Wires of very small diameter, between 10 and 100~$\mu$m, are placed between two metallic plates \ced{a few millimetres apart}. The wires are at a high voltage of a few kilovolts, which results in a very high electric field close to the wire surface. The ionization electrons \ced{move towards the thin wires, and, in the strong fields close to the wires}, the electrons are accelerated to energies above the ionization energy of the gas, which results in secondary electrons and as a consequence an electron avalanche. Gas gains of $10^4$--$10^5$ are typically used, which makes the wire chambers perfectly sensitive to single tracks. In this basic application, the position of the track is therefore given by the position of the wire that carries a signal, so we have a one-dimensional positioning device.

One has to keep in mind that the signal in the wire is not due to the electrons entering wire; rather, the signal is induced while the electrons are moving towards the wire and the ions are moving away from it. Once all charges arrive at the electrode, the signal is terminated. The signals in detectors based on ionization are therefore {\it induced} on the readout electrodes by the {\it movement} of the charges. This means that we find signals not only on electrodes that receive charges but also on other electrodes in the detector. For the wire chamber one can therefore segment the metal plates (cathodes) into strips in order to find the second coordinate of the track along the wire direction. In many applications, one does not even read out the wire signals but instead one segments the cathode planes into square or rectangular pads to get the full two-dimensional information from the cathode pad readout. The position resolution is in this case not limited by the pad size. If one uses pad dimensions of the order of the cathode-to-wire distance, one finds signals on a few neighbouring pads, and, by using centre-of-gravity interpolation, one can determine the track position, which is only 1/10 to 1/100 of the pad size. Position resolution down to 50~$\mu$m and rate capabilities of \ced{hundreds of kHz of particles per cm$^2$ per second}\aq{Is this use of kHz here correct?} can be achieved with these devices.

Another way to achieve position resolution that is far smaller than the wire \ced{separation}\aq{Is ``separation'' correct rather than ``distance''?} is the so-called drift chamber. One determines the time when the particle passes the detector by an external device, which can be a scintillator or the accelerator clock in a collider experiment, and one uses the arrival time of the ionization electrons at the wire as the measure of the distance between the track and the wire. The ATLAS muons system, for instance, uses tubes of 15~mm radius with a central wire, and the measurement of the drift time determines the track position to 80~$\mu$m precision.

The choice of the gas for a given gas detector is dominated by the transport properties of electrons and ions in gases, because these determine the signal and timing characteristics. In order to avoid the ionization electrons getting lost on their way to the readout wires, one can use only gases with very small electronegativity. The main
component of detector gases are therefore the noble gases like argon or neon. Other admixtures like hydrocarbons (methane, isobutane) or CO$_2$ are also needed in order to ``tune'' the gas transport properties and to ensure operational stability. Since hydrocarbons were shown to cause severe chamber ageing effects at high rates, the LHC detectors use almost exclusively argon, neon and xenon together with CO$_2$ for all wire chambers.

Typical drift velocities of electrons are in the range of 5--10~cm/$\mu\mathrm{s}$. The velocity of the ions that are produced in the electron avalanche at the wire and are moving back to the cathodes is about 1000--5000 times smaller than the electron velocity. The movement of these ions produced long signal tails in wire chambers, which have to be properly removed by dedicated filter electronics.

During the past 10--15 years a very large variety of new gas detectors have entered particle physics instrumentation, the so-called micropattern gas detectors like the GEM (gas electron multiplier) or the MICROMEGA (micro mesh gas detector). In these detectors the high fields for electron multiplication are produced by micropattern structures that are realized with photolithographic methods. Their main advantages are rate capabilities far in excess of those achievable in wire chambers, low material budget construction and semi-industrial production possibilities.

\subsection{Solid-state detectors}

In gaseous detectors, a charged particle liberates electrons from the atoms, which are freely bouncing between the gas atoms. An applied electric field makes the electrons and ions move, which induces signals on the metal readout electrodes. For individual gas atoms, the electron energy levels are discrete.

In solids (crystals), the electron energy levels are in ``bands''. Inner-shell electrons, in the lower energy bands, are closely bound to the individual atoms and always stay with ``their'' atoms. However, in a crystal there are energy bands that are still bound states of the crystal, but they belong to the entire crystal. Electrons in these bands and the holes in the lower band  can move freely around the crystal, if an electric field is applied. The lowest of these bands is called the ``conduction band''.

If the conduction band is filled, the crystal is a conductor. If the conduction band is empty and ``far away'' from the last filled band, the valence band, the crystal is an insulator. If the conduction band is empty but the distance to the valence band is small, the crystal is called a semiconductor.

The energy gap between the valence band and the conduction band is called the band gap $E_g$. The band gaps of diamond, silicon and germanium are 5.5, 1.12 and 0.66~eV, respectively. If an electron in the valence band gains energy by some process, it can be excited into the conduction band and a hole in the valence band is left behind. Such a process can be the passage of a charged particle, but also thermal excitation with a probability proportional to $\exp(-E_g/kT)$. The number of electrons in the conduction band therefore increases with temperature, i.e.\ the conductivity of a semiconductor increases with temperature.

It is possible to treat electrons in the conduction band and holes in the valence band similar to free particles, but with an effective mass different from elementary electrons not embedded in the lattice. This mass is furthermore dependent on other parameters such as the direction of movement with respect to the crystal axis.
If we want to use a semiconductor as a detector for charged particles, the number of charge carriers in the conduction band due to thermal excitation must be smaller than the number of charge carriers in the conduction band produced by the passage of a charged particle.  Diamond can be used for particle detection at room temperature; silicon and germanium must be cooled, or the free charge carriers must be eliminated by other tricks like ``doping''.

The average energy to produce an electron--hole pair for diamond, silicon and germanium, respectively, is 13, 3.6 and 2.9~eV. Compared to gas detectors, the density of a solid is about a factor of~1000 larger than that of a gas, and the energy to produce an electron--hole pair for silicon, for example, is a factor~7 smaller than the energy to produce an electron--ion pair in argon. The number of primary charges in a silicon detector is therefore about $10^4$ times larger than in a gas and, as a result,  solid-state detectors do not need internal amplification. While, in gaseous detectors, the velocities of electrons and ions differ by a factor of~1000, the velocities of electrons and holes in many semiconductor detectors are quite similar, which results in very short signals of a few tens of nanosecond length.

The diamond detector works like a solid-state ionization chamber. One places diamond of a few hundred micrometres thickness between two metal electrodes and applies an electric field. The very large electron and hole mobilities of diamond result in very fast and short signals, so, in addition to tracking application, the diamond detectors are used as precision timing devices.

Silicon is the most widely used semiconductor material for particle detection. A high-energy particle produces around 33\,000 electron--hole pairs in 300~$\mu$m of silicon. At room temperature there are, however, $1.45 \times 10^{10}$ electron--hole pairs per cm$^3$. To apply silicon as a particle detector at room temperature, one therefore has to use the technique of ``doping''. Doping silicon with \ced{arsenic} makes it an n-type conductor (more electrons than holes); doping silicon with boron makes it a p-type conductor (more holes that electrons). Putting an n-type and p-type conductor in contact realizes a diode.

At a p--n junction the charges are depleted and a zone free of charge carriers is established. By applying a voltage, the depletion zone can be extended to the entire diode, which results in a highly insulating layer. An ionizing particle produces free charge carriers in the diode, which  drift in the electric field and therefore induce an electrical signal on the metal electrodes. As silicon is the most commonly used material in the electronics industry, it has one big advantage with respect to other materials, namely highly developed technology.

Strip detectors are a very common application, where the detector is segmented into strips of a few 50--150~$\mu$m pitch and the signals are read out on the ends by wire bonding the strips to the readout electronics. The other coordinate can then \ced{be determined, either by another strip detector with perpendicular orientation, or by implementing perpendicular strips} on the same wafer. This technology is widely used at the LHC, and the CMS tracker uses 445~m$^2$ of silicon detectors.

In the very-high-multiplicity region close to the collision point, a geometry of crossed strips results in too many ``ghost'' tracks, and one has to use detectors with a chessboard geometry, so-called pixel detectors, in this region. The major complication is the fact that each of the chessboard pixels must be connected to a separate readout electronics channel. This is achieved by building the readout electronics wafer in the same geometry as the pixel layout and soldering (bump bonding) each of the pixels to its respective amplifier. Pixel systems in excess of 100 million channels are successfully operating at the LHC.

A clear goal of current solid-state detector development is the possibility of integration of the detection element and the readout electronics into a monolithic device.

\section{Calorimetry}

The energy measurement of charged particles by completely absorbing (``stopping'') them is called calorimetry. Electromagnetic (EM) calorimeters measure the energy of electrons and photons. Hadron calorimeters measure the energy of charged and neutral hadrons.

\subsection{Electromagnetic calorimeters}

As discussed above, high-energy electrons suffer significant bremsstrahlung owing to their small mass. The interplay of bremsstrahlung and pair production will develop a single electron or photon into a shower of electrons and positrons. The energy of these shower particles decreases exponentially until all of them are stopped due to ionization loss. The total amount of ionization produced by the electrons and positrons is then a measure of the particle energy. The characteristic length scale of this shower process is called the radiation length $X_0$, and in order to fully absorb a photon or electron one typically uses a thickness of about $25\,X_0$. One example of such an EM calorimeter at the LHC is the crystal calorimeter of CMS, which uses PbW$_4$ crystals. The radiation length $X_0$ of this crystal is 9~mm, so with a length of 22~cm one can fully absorb the high-energy electron and photon showers. In these crystals the light produced by the shower particles is used as the measure of the energy.

Liquid noble gases are the other prominent materials used for EM calorimetry. In these devices, the total amount of ionization is used as a measure of the energy. The NA48 experiment uses a homogeneous calorimeter of liquid krypton, which has a radiation length of 4.7~cm. Liquid argon has a radiation length of 14~cm, so one would need a depth of 350~cm to fully absorb the EM showers. Since this is not \ced{practicable}, one interleaves the argon with absorber material of smaller radiation length, such as lead, to allow a more compact design of the calorimeter. Such an alternating assembly of absorber material and active detector material is called a sampling calorimeter. Although the energy resolution of such a device is worse compared to a homogeneous calorimeter, for many applications  it is good enough. The ATLAS experiment uses such a liquid argon sampling calorimeter. Other calorimeter types use plastic scintillators interleaved with absorber materials.

The energy resolution of calorimeters improves as $1/\sqrt{E}$ where $E$ is the particle energy. This means that the energy measurement becomes ``easier'' at high-energy colliders. For homogeneous EM calorimeters, energy resolutions of $\sigma_E/E = 1\%/\sqrt{E\,(\mbox{GeV})}$ are achieved; typical resolutions of sampling calorimeters are in the range of $\sigma_E/E = (10\mbox{--}20\%)/\sqrt{E\,(\mbox{GeV})}$.

\subsection{Hadron calorimeters}

While only electrons and photons have small enough masses to produce significant EM bremsstrahlung, there is a similar ``strong-interaction bremsstrahlung effect'' for hadrons. High-energy hadrons radiate pions in the vicinity of a nucleus, and a cascade of these pions develops, which also fully absorbs the incident hadron, and the total ionization loss of this cascade is used to measure the particle energy. The length scale of this shower development is the so-called hadronic interaction length $\lambda$, which is significantly larger than the radiation length $X_0$. For iron the radiation length $X_0$ is 1.7~cm, whereas the hadronic interaction length $\lambda$ is 17~cm. Hadron calorimeters are therefore significantly larger and heavier than EM calorimeters. The energy resolution of hadron calorimeters is typically worse than that of EM calorimeters because of the more complex shower processes. About 50\% of the energy ends up in pions, 20\% ends up in nuclear excitation and 30\% goes into slow neutrons, which are usually not detected. A fraction of the produced pions consists of $\pi_0$, which instantly decay into two photons, which in turn start an EM cascade. The relative fluctuations of all these processes will result in a larger fluctuation of the calorimeter signal and therefore reduced resolution. Hadron calorimeters are also typically realized as sampling calorimeters with lead or steel plates interleaved with scintillators or liquid noble gases. Energy resolutions of $\sigma_E/E = (50\mbox{--}100\%)/\sqrt{E\,(\mbox{GeV})}$ are typical.

\section{Particle identification}

By measuring the trajectory of a particle in a magnetic field, one measures the particle's momentum, so in order to determine the particle type, i.e.\ the particle's mass, one needs an additional measurement. Electrons, positrons and photons can be identified by electromagnetic calorimetry, and muons can be identified by the fact that they traverse large amounts of material without being absorbed.  To distinguish between protons, kaons and pions is a slightly more subtle affair, and it is typically achieved by measuring the particle's velocity in addition to the momentum.

For kinetic energies that are not too far from the rest mass of the particle, the velocity is not yet too close to the speed of light, such that one can measure the velocity by time of flight. With precision timing detectors like scintillators or resistive plate chambers, time resolutions of less than 100~ps are being achieved. For a time-of-flight distance of 1~m, this allows kaon/pion separation up to 1.5~GeV/$c$, and proton/pion separation up to about 3~GeV/$c$.

The energy loss of a particle also measures its velocity, so particle identification up to tens of GeV for pions and protons can be achieved. In gas detectors with pad readout and charge interpolation, the signal pulse height is measured for centre-of-gravity interpolation in view of precision tracking. Since the pulse height is a measure of the energy loss, it can in addition be used for particle identification. Time projection chambers are the best examples of combined tracking and particle identification detectors.

For larger velocities, one can use the measurement of the Cerenkov angle to find the particle velocity. This radiation is emitted at a characteristic angle that is uniquely related to the particle velocity. Using short radiators this angle can be determined simply by measuring the radius of the circle produced by the photons in a plane at a given distance from the radiator. Another technique uses a spherical mirror to project the photons emitted along a longer path onto a plane that also forms a circle. Detectors of this type are called ring imaging Cerenkov detectors (RICH). Since only a ``handful'' of photons are emitted over typical radiator thicknesses, very efficient photon detectors are the key ingredient to Cerenkov detectors. Using very long gas radiators with very small refractive index, kaon/pion separation up to momenta of 200~GeV/$c$ has been achieved.

\section{Signal readout}

Many different techniques to make particle tracks visible were developed in the last century. The cloud chamber, the bubble chamber and the photographic emulsion were taking actual pictures of the particle tracks. Nowadays we have highly integrated electronic detectors that allow high particle rates to be processed with high precision. Whereas bubble chambers were almost unbeatable in terms of position resolution (down to a few micrometres) and the ability to investigate very complex decay processes, these detectors were only able to record a few events per second, which is not suitable for modern high-rate experiments. The LHC produces $10^9$ proton--proton collisions per second, of which, for example, 100 produce W~bosons that decay into leptons, 10 produce a top quark pair and 0.1 produce a hypothetical Higgs particle of 100~GeV. Only around 100 of the $10^9$ events per second can be written to tape, which still results in petabytes of data per year to be analysed. The techniques to reduce the rate from $10^9$ to 100~Hz by selecting only the ``interesting'' events is the realm of the so-called trigger and data acquisition. With a bunch crossing time of 25~ns, the particles produced in one collision have not even reached the outer perimeter of the detector when the next collision is already taking place. The synchronization of the data belonging to one single collision is therefore another very challenging task. In order to become familiar with the techniques and vocabulary of trigger and data acquisition, we discuss a few examples.

If, for example, we want to measure temperature, we can use the internal clock of a PC to periodically trigger the measurement. If, on the other hand, we want to measure the energy spectrum of the beta-decay electrons of a radioactive nucleus, we need to use the signal itself to trigger the readout. We can split the detector signal caused by the beta electron and use one path to apply a threshold to the signal, which produces a ``logic'' pulse that can ``trigger'' the measurement of the pulse height in the second path. Until this trigger signal is produced, one has to ``store'' the signal somewhere, which is done in the simplest application by a long cable where the signal can propagate.

If we measure the beta electrons, we cannot distinguish the signals from cosmic particles that are traversing the detector. By building a box around our detector that is made from scintillator, for example, we can determine whether a cosmic particle has entered the detector or whether it was a genuine beta-decay electron. Triggering the readout on the condition of a detector signal in coincidence with the absence of a signal in the scintillator box, we can therefore arrive at a pure beta spectrum sample.

Another example of a simple ``trigger'' logic is the measurement of the muon lifetime with a stack of three scintillators. Many of the cosmic muons will \ced{pass through} all three scintillators, but some of them will have lower energy such that they traverse the first one and get stuck in the central one. After a certain time the muon will decay and the decay electron produces a signal in the central and the bottom scintillators. By starting a clock with a signal condition of 1\,AND\,2\,ANDNOT\,3 and stopping the clock with NOT\,1\,AND\,2\,AND\,3, one can measure the lifetime of the muons.

At the LHC experiment some typical trigger signals are high-energy events transverse to the proton beam direction, which signify interesting high-energy parton collisions. High-energy clusters in the calorimeters or high-energy muons are therefore typical trigger signals, which start the detector readout and ship the data to dedicated processing units for further selection refinement.

In order to cope with high rates, one has to find appropriate ways to deal with the ``processing'' time, i.e.\ the time while the electronics is busy with reading out the data. This we discuss in the following. First we assume a temperature sensor connected to a PC. The PC has an internal clock, which can be used to periodically trigger the temperature measurement and write the values to disk. The measurement and data storage will take a certain time $\tau$, so this ``deadtime'' limits the maximum acquisition rate. For a deadtime $\tau = 1$~ms, we have a maximum acquisition rate of $f=1/\tau=1$~kHz.

For the example of the beta spectrum measurement, we are faced with the fact that the events are completely random and it can happen that another beta decay takes place while the acquisition of the previous one is still ongoing. In order to avoid triggering the readout while the acquisition of the previous event is still ongoing, one has to introduce a so-called ``busy logic'', which blocks the trigger while the readout is ongoing. Because the time between events typically follows an exponential distribution, there will always be events lost even if the acquisition time is smaller than the average rate of events. In order to collect 99\% of the events, one has to overdesign the readout system with a deadtime of only 10\% of the average time between events. To avoid this problem, one uses a so-called FIFO (first-in first-out) buffer in the data stream. This buffer receives as input the randomly arriving data and stores them in a queue. The readout of the buffer happens at constant rate, so by properly choosing the depth of the buffer and the readout rate, it is possible to accept all data without loss, even for readout rates close to the average event rate. This transformation from random input to clocked output is call ``de-randomization''.

In order to avoid ``storing'' the signals in long cables, one can also replace them by FIFOs. At colliders, where the bunch crossing comes in regular intervals, the data are stored in so-called front-end pipelines, which sample the signals at the bunch crossing rate and store them until a trigger decision arrives.

The event selection is typically performed at several levels of increasing refinement.
The fast trigger decisions in the LHC experiments are performed by specialized hardware on or close to the detector. After a coarse events selection, the rates are typically low enough to allow a more refined selection using dedicated computer farms that do more sophisticated analysis of the events. The increasing computing power, however, drives the concepts of trigger and \ced{data acquisition} into quite new directions. The concepts for some future high-energy experiments foresee so-called ``asynchronous'' data-driven readout concepts, where the signal of each detector element receives a time stamp and is then shipped to a computer farm where the event synchronization and events selection is carried out purely by software algorithms.


\end{document}